\documentclass[12pt]{amsart}
\usepackage{mathrsfs}
\usepackage{mathtools} 
\mathtoolsset{showonlyrefs}
\usepackage{algorithm,algorithmic}
 
\usepackage{boites}

\usepackage{fullpage}

\usepackage[dvipdfmx]{graphicx}
 
\usepackage{xcolor}
\usepackage{times}
\usepackage{bm}
\usepackage{natbib}
\usepackage{amsmath,amssymb,bbm}
\usepackage{enumerate}
\usepackage{xspace}
\usepackage{array}
\newcolumntype{M}[1]{>{\centering\arraybackslash}m{#1}}

\newcommand{\Eppos}{\Ep_{\mathrm{pos}}}
\newcommand{\Covpos}{\mathrm{Cov}_{\mathrm{pos}}}

\newcommand{\Kpos}{{\mathcal{K}_{\mathrm{pos}}}}
\newcommand{\KposABBBw}{{\mathcal{K}^{(1,3),w}_{\mathrm{pos}}}}

\newcommand{\Epposl}{\Ep^\lambda_{\mathrm{pos}}}
\newcommand{\Covposl}{\mathrm{Cov}^{\lambda}_{\mathrm{pos}}}

\newcommand{\Kposl}{{\mathcal{K}^{\lambda}_{\mathrm{pos}}}}

\newcommand{\Ep}{\mathbf{E}}
\newcommand{\EpX}{\Ep_{X}}
\newcommand{\EpXi}{\Ep_{X_1}}

\usepackage{natbib}
\usepackage{appendix}
 
\usepackage{enumitem}

\begin{document}


%

%

\title{Bias correction of posterior means using MCMC outputs}
\author{Yukito Iba}
\address{The Institute of Statistical Mathematics,
10-3 Midori cho, Tachikawa City, Tokyo 190-8562, Japan.} 
\email{iba@ism.ac.jp}
 
\begin{abstract}
We propose algorithms for addressing the bias of the posterior mean when used as an estimator of parameters. These algorithms build upon the recently proposed Bayesian infinitesimal jackknife approximation (\cite{Bayesian_IJK}) and can be implemented using the posterior covariance and third-order combined cumulants easily calculated from MCMC outputs. Two algorithms are introduced: The first algorithm utilises the output of a single-run MCMC with the original likelihood and prior to estimate the bias. A notable feature of the algorithm is that its ability to estimate {\it definitional bias} (\cite{Efron_2015}), which is crucial for Bayesian estimators. The second algorithm is designed for high-dimensional and sparse data settings, where ``quasi-prior'' for bias correction is introduced. The quasi-prior is iteratively refined using the output of the first algorithm as a measure of the residual bias at each step. These algorithms have been successfully implemented and tested for parameter estimation in the Weibull distribution and logistic regression in moderately high-dimensional settings.
\end{abstract}

\maketitle
 

\section{Introduction}
\label{sec:introduction}

This paper presents algorithms for estimating and correcting the bias of the posterior mean when used as an estimator of parameters. We discuss two algorithms, both of which are based on posterior covariance and third-order combined cumulants.

The first algorithm utilizes the output of a single run of MCMC with the original likelihood and prior. A simple formula can be applied to any likelihood and prior, and the estimator is automatically computed from the posterior samples; no model-specific analytical calculation is required.

The second algorithm is designed for high-dimensional and sparse data settings, where a “quasi-prior” for bias correction is introduced. The quasi-prior is improved iteratively using the output of the first algorithm as a measure of the residual bias at each step. Typically, after several MCMC runs, the bias correction by the first algorithm yields satisfactory results; it is important to note that we do not need to entirely remove the bias through this iteration. Experiments show that the second algorithm is effective, for example, in models with 20 or 60 parameters.

These algorithms heavily rely on the recently proposed idea of the Bayesian infinitesimal jackknife approximation (Bayesian IJK, \cite{Bayesian_IJK}). While frequentist covariance is represented by the posterior covariance in \cite{Bayesian_IJK}, our algorithm estimates bias using posterior cumulants; we also refer to a related idea suggested in \cite{Efron_2015}. The proposed algorithms, however, accommodate the following novel features:
\begin{enumerate}[label=(\alph*)]
    \item The {\it definitional bias} (\cite{Efron_2015}) is included in the proposed algorithms; this is essential for the Bayesian estimators. 
 
    \item Iterative tuning of the quasi-prior is introduced in the second algorithm. This remarkably improves the performance of the algorithm in high-dimensional and sparse-data settings.
\end{enumerate}

Details of (a) and (b) are discussed in the remainder of the paper. Here we provide a simple example of the definitional bias. Consider a binomial likelihood with a beta prior $Beta(\alpha,\beta)$ on the success probability $q$, and the posterior mean 
$
\Eppos[q]= 
{(X+\alpha)}/{(n+\alpha+\beta)}
$
as an estimator of $q$, where $n$ and $X$ is the number of the trials and successes, respectively. In this context, the definitional bias $b_0$ is expressed as
$
b_0
= 
{(n\EpX[X]+\alpha)}/{(n+\alpha+\beta)}
-q_0
= 
{(nq_0+\alpha)}/{(n+\alpha+\beta)}
-q_0,
$
where $\EpX[X]$ denotes the population average of $X$. Since the estimator $\Eppos[q]$ is linear in $X$, the bias caused by the non-linearity of the estimator is zero. Definitional bias $b_0$ is often disregarded in discussions based on the von Mises expansion or nonparametric bootstrap; in many theoretical  analyses, the ``bias'' of an estimator is defined as the deviation from $\theta_0+b_0$ rather than from the population value $\theta_0$ of the paramete. However, numerical experiments indicate that definitional bias $b_0$ can be significantly large for posterior mean estimators even when seemingly noninformative priors are used. This contrasts with the case of the maximum likelihood estimator, where $b_0$ is always 0. In this study, we introduce a simple method to estimate $b_0$ using the posterior covariance. The proposed estimator may fail in sparse data settings, but this issue is addressed by the use of the quasi-prior and its iterative improvement, as employed in the second algorithm.  

The rest of the paper is organized as follows: In Sec.~\ref{sec:proposed_method}, we introduce the proposed algorithms. In sections \ref{sec:example_Weibull} and \ref{sec:example_logistic}, illustrative examples are presented for these algorithms. In Sec.~\ref{sec:derivation}, the proposed algorithms are derived from the jackknife approximation to the bias. Sec.~\ref{sec:derivation} provides a summary and conclusion. Appendix~\ref{app:von_Mises} defines the bias and definitional bias by means of the von Mises expansion and Appendix~\ref{app:jack} derives the jackknife bias correction formula. 

\section{Proposed method}
\label{sec:proposed_method}

\subsection*{Settings and notation}

Let us denote a sample of size $n$ from the population $G$ as $X^{n}=(X_1,\ldots, X_n)$.
The posterior distribution is defined by
\begin{align}
    p(\theta \mid X^{n} ) = 
    \frac{ \exp\{\sum_{i=1}^{n} \ell(X_{i}; \theta)\}p(\theta) }
    {\int \exp\{\sum_{i=1}^{n} \ell(X_{i};\theta')\}p(\theta') d\theta'},
    \label{eq:pos}
\end{align}
where $p(\theta)$ is a prior density on $\theta=(\theta_1,\cdots, \theta_K)$ and $\ell(x;\theta)=\log p(x |\theta)$ is log-likelihood of the model. 

Our objective is to correct the frequentist bias in the posterior mean defined by
\begin{align}
\Eppos[A]  = \int A(\theta)p(\theta \mid X^n) d\theta
\end{align} 
of given statistics $A(\theta)$. We also define the posterior covariance and a third-order combined posterior cumulant of the arbitrary statistics $A(\theta)$, $B(\theta)$, and $C(\theta)$ as
\begin{align}
&\Covpos[A(\theta),B(\theta)]  =
\Eppos[(A(\theta)-\Eppos[A(\theta)])( B(\theta)-\Eppos[B(\theta)])],
\\
& \Kpos[A(\theta),B(\theta), C(\theta)]  =
\Eppos[(A(\theta)-\Eppos[A(\theta)])(B(\theta)-\Eppos[B(\theta)])(C(\theta)-\Eppos[C(\theta)]]).
 \label{eq:poscum_3}
\end{align} 
\\  

\subsection*{Algorithm 1}

In the algorithm 1, the bias $b(A;n)$ is estimated using the formula
\begin{align}
\hat{b}(A;n)=  - \sum_{i=1}^n \Covpos[A(\theta), \ell(X_{i}; \theta)]+\frac{1}{2} \sum_{i=1}^n \Kpos[A(\theta), \ell(X_{i}; \theta), \ell(X_{i}; \theta)].
\label{eq:b_cum}
\end{align}
In practice, the posterior covariance and cumulants in the above formula are estimated from posterior samples obtained from a single MCMC run. The resultant algorithm is summarized as follows:

 \begin{breakbox}
 \begin{algorithmic}
 \REQUIRE Observations $X^n=(X_{1},\ldots,X_{n})$ and the number $M$ of posterior samples.
 \ENSURE  An estimate of the bias $b$ of the estimator $\Eppos[A]$\\ 
 
 \textit{Step 1} :
  \STATE Sample\,\,
  $\theta^{(1)},\ldots,\theta^{(M)}$ from the posterior $p(\theta \mid X^n).$\\ 
 
 \textit{Step 2} :
  \STATE Calculate for $i=1,\ldots,n$,
  \begin{align*}
  & \bar{\ell}_i=\frac{1}{M}\sum_{j=1}^{M} \ell(X_i;\theta^{(j)}), 
  \,\,\, 
  \bar{A}=\frac{1}{M}\sum_{j=1}^{M} A(\theta^{(j)}). \\
  & D_{1,i}= \frac{1}{M}\sum_{j=1}^{M} (\ell(X_i;\theta^{(j)})-\bar{\ell}_i)(A(\theta^{(j)})-\bar{A}), 
  \,\,\,
  D_{2,ii}= \frac{1}{M}\sum_{j=1}^{M} (\ell(X_i;\theta^{(j)})-\bar{\ell}_i)^2 (A(\theta^{(j)})-\bar{A}).\\
  \end{align*}
  Calculate
  \begin{align*}
  \hat{b}(A;n)=-\sum_{i=1}^n D_{1,i} + \frac{1}{2}\sum_{i=1}^n D_{2,ii}  
  \end{align*}  
  \\
  
 \RETURN $\hat{b}(A;n)$ as an estimate the bias of the posterior mean $\Eppos[A]$
 \end{algorithmic}
 \end{breakbox}

\subsection*{Algorithm 2}

In the algorithm 2, simultaneous bias correction of all parameters is essential. For simplicity, here we restrict to the case of simultaneous bias correction of the original parameters $(\theta,\cdots,\theta_K)$, while general cases including an arbitrary statistics $A(\theta)$ may be dealt with a change of the parameter.  

As a basis of the algorithm,  we introduce a modified posterior distribution defined by
\begin{align}
    p_\lambda(\theta \mid X^{n} ) = 
    \frac{ \exp\{\sum_{i=1}^{n} \ell(X_{i}; \theta)-\sum_{k=1}^K \lambda_k \theta_k \}p(\theta) }
    {\int \exp\{\sum_{i=1}^{n} \ell(X_{i};\theta')-\sum_{k=1}^K \lambda_k \theta'_k\}p(\theta') d\theta'},
    \label{eq:pos_lambda}
\end{align}
where $\lambda=(\lambda_1,\cdots,\lambda_K)$ is a vector of constants. An average $\Epposl[\,\,\,]$, covariance $\Covposl[\,\,\,]$, and third-order combined cumulant $\Kposl[\,\,\,]$ with respect to the distribution \eqref{eq:pos_lambda} are defined in a similar manner that $\Eppos[\,\,\,]$, $\Covpos[\,\,\,]$, and $\Kpos[\,\,\,]$ is defined, respectively. Using them, we define
\begin{align}
    & \hat{b}(\theta_k;n;\lambda)=  - \sum_{i=1}^n \Covposl[\theta_k, \ell(X_{i}; \theta)]+\frac{1}{2} \sum_{i=1}^n \Kposl[\theta_k, \ell(X_{i}; \theta), \ell(X_{i}; \theta)] \label{eq:b_b0_C},
\shortintertext{and}
    & C(\theta_k,\theta_{k^\prime};n;\lambda)=\Covposl[\theta_k, \theta_{k^\prime}].
\end{align}
Hereafter, the vector whose components is $\hat{b}(\theta_k;n;\lambda)$ is denoted as $\hat{b}(\theta;n;\lambda)$, while the matrix whose ($k,k^\prime$) component is $C(\theta_k,\theta_{k^\prime};n;\lambda)$ is expressed as $C(\theta;n;\lambda)$. 

The proposed procedure is an iterative improvement of the constants $\lambda=(\lambda_1,\cdots,\lambda_K)$, which intends to reduce the bias $\hat{b}(\theta;n;\lambda)$ estimated in each step. Let us denote the value of $\lambda$ at step $m$ as $\lambda^{(m)}$ and define $\triangle\lambda^{(m)}$ as a vector that solves a linear equation 
\begin{align}
    C(\theta;n;\lambda^{(m)}) \triangle\lambda^{(m)} = 
    \hat{b}(\theta;n;\lambda^{(m)}). 
    \label{eq:iter_1}
\end{align} 
Using $\triangle\lambda^{(m)}$, the value of $\lambda^{(m)}$ is updated as
\begin{align}
    \lambda^{(m+1)}=\lambda^{(m)}+\delta \, \triangle\lambda^{(m)}, 
    \label{eq:iter_2}
\end{align}   
where $0< \delta \leq 1$ is a constant in the algorithm, the value of which is chosen to prevent oscillation of the value of $\lambda$. Since $C(\theta;n;\lambda^{(m)})$ is a  matrix that express the posterior covariance between parameters, the above equation has a unique solution, unless parameters are not identifiable or extremely correlated. 

An essential point is that {\it we need not entirely remove the bias} by the iteration procedure. Our strategy is to reduce the magnitude of the bias towards the range where the approximation \eqref{eq:b_b0_C} is sufficiently accurate, then estimate it using \eqref{eq:b_b0_C}.\\

The resultant algorithm is summarized as follows:
 \begin{breakbox}
 \begin{algorithmic}
 \REQUIRE Observations $X^n=(X_{1},\ldots,X_{n})$, the number $M$ of posterior samples, the constant $\delta$, and the number $L$ of the iteration. 
 \ENSURE  A bias-corrected estimate 
 of the parameters $\theta=(\theta_1,\cdots,\theta_K)$.\\ 
 
 \textit{Step 1} : 
  \STATE Set $l=1$. Set $\lambda=0$.

  \textit{Step 2} : 
  \REPEAT
  \STATE Sample\,\,
  $\theta^{(1)},\ldots,\theta^{(M)}$ from the posterior $p_\lambda(\theta \mid X^n)$ defined by \eqref{eq:pos_lambda}.\\ 
  \STATE Calculate \, $\hat{b}(\theta;n;\lambda)$ \, by the algorithm 1.
  \STATE Calculate \, $\bar{\theta}=\frac{1}{M}\sum_{j=1}^{M} \theta^{(j)}$. 
  \STATE Calculate \, $C(\theta_k,\theta_{k^\prime};n;\lambda) = \frac{1}{M}\sum_{j=1}^{M} (\mathstrut \theta_k^{(j)}-\bar{\theta}_k)(\mathstrut \theta_{k'}^{(j)}-\bar{\theta}_{k'})$ \, for $k,k'= 1, \ldots, K$ .\\
  \STATE Define \, $C(\theta;n;\lambda)$ as a matrix whose $(k,k^\prime)$ component is $C(\theta_k,\theta_{k^\prime};n;\lambda)$.
  \STATE Calculate \, $\triangle\lambda$ \, by solving \,
  $C(\theta;n;\lambda) \triangle\lambda = \hat{b}(\theta;n;\lambda)$. 
  \STATE Update \, $\lambda=\lambda+\delta \, \triangle\lambda$. 
  \STATE Set $l=l+1$.
  \UNTIL $l=L$.
 
 \textit{Step 3} :
  \STATE Sample\,\,
  $\theta^{(1)},\ldots,\theta^{(M)}$ from the posterior $p_\lambda(\theta \mid X^n)$ defined by \eqref{eq:pos_lambda}.\\ 
  \STATE Calculate \,\, $\hat{b}(\theta;n;\lambda)$ \, by the algorithm 1.
  \STATE Calculate \,\,
   $\hat{\theta}(\lambda)=\frac{1}{M}\sum_{j=1}^{M} \theta^{(j)}$.
  \\
  
 \RETURN 
 $\hat{\theta}(\lambda)+\hat{b}(\theta;n;\lambda)$ as an estimate of the parameters $\theta=(\theta_1,\cdots,\theta_K)$.
 \end{algorithmic}
 \end{breakbox}

\subsection*{Definition of $\hat{b}_0$ and $\hat{b}_2$}

Since the first and second term in \eqref{eq:b_cum} have a considerably different nature, contribution of each term is checked separately in the following numerical experiments. Hereafter, we denote the first and second term in \eqref{eq:b_cum} as
\begin{align}
  \hat{b}_0(A;n)= & - \sum_{i=1}^n \Covpos[A(\theta), \ell(X_{i}; \theta)], \label{eq:b0} \\
  \hat{b}_2(A;n)= & \frac{1}{2} \sum_{i=1}^n \Kpos[A(\theta), \ell(X_{i}; \theta), \ell(X_{i}; \theta)], \label{eq:b2}\\
\end{align}  
which can be abbreviated as $\hat{b}_0(A)$ and $\hat{b}_2(A)$, or even $\hat{b}_0$ and $\hat{b}_2$, respectively. By an abuse of notation, the corresponding components of $\hat{b}(A;n;\lambda)$ are also expressed by the same symbols, such as $\hat{b}_0$ and $\hat{b}_2$.

In appendix \ref{app:von_Mises}, we define two components $b_0(A;n)$ and $b_2(A;n)$ of the bias based on the von Mises expansion; $\hat{b}_0(A;n)$ and $\hat{b}_2(A;n)$ are considered as an estimator of $b_0(A;n)$ and $b_2(A;n)$, respectively.  

\section{Example: Weibull fitting}
\label{sec:example_Weibull}

Here we test the algorithm 1 for the parameter estimation of the Weibull distribution, the probability density of which is given by
\begin{align}
    p^{wb}(x \mid  \lambda, \gamma)=\frac{\gamma}{\lambda}\left (\frac{x}{\lambda} \right )^{\gamma-1}\exp \left [-\left ( \frac{x}{\lambda} \right)^\gamma \right ], \,\,\, x \geq 0.
    \label{eq:weibull_PDF}
\end{align} 
This likelihood is not an exponential family in the shape parameter $\gamma$. An improper prior uniform on $[0,\infty)$ is assumed for each of the parameters $\lambda$ and $\gamma$. Experiments below are performed with 5000 sets of artificial data of sample size $n=30$; they are generated from the Weibull distribution with $\lambda=1.0$ and $\gamma=0.8$. 

\begin{figure}[th]
    \centering
    \includegraphics[width=60mm]{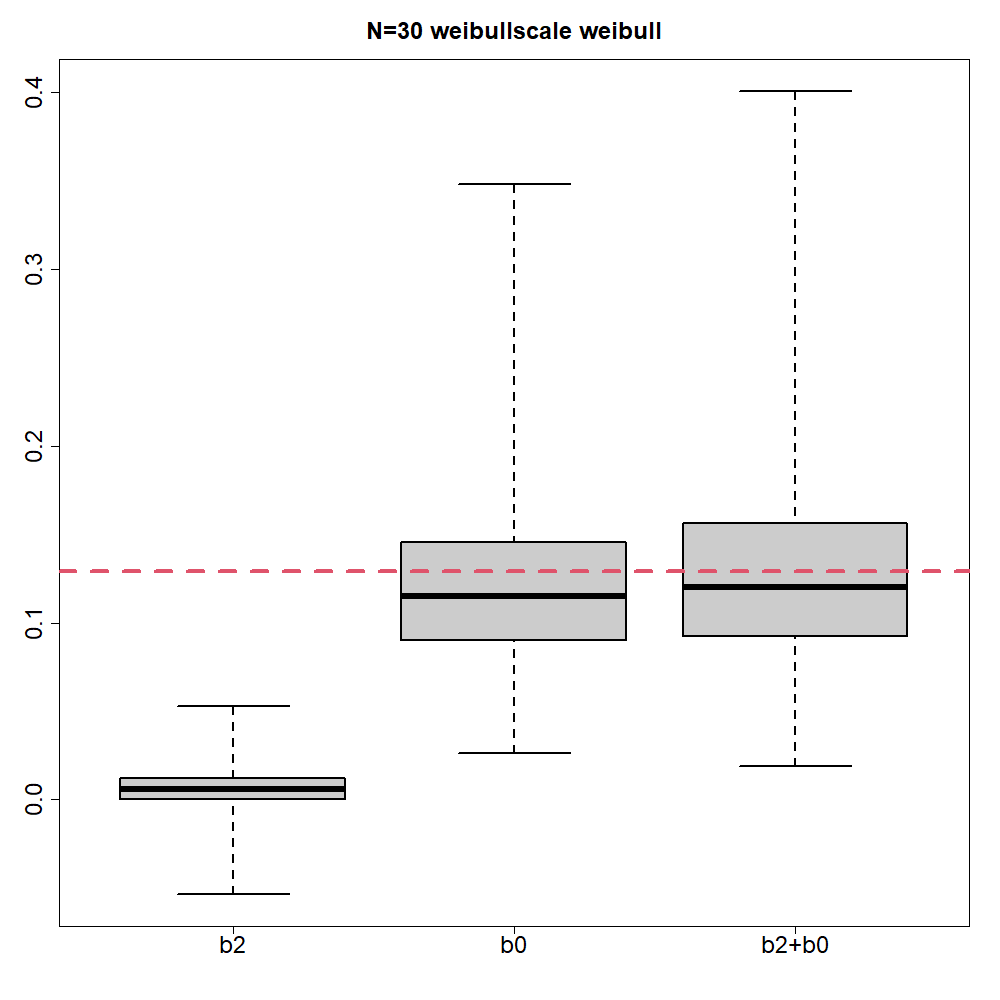}
    \hspace{0.5cm}
    \includegraphics[width=60mm]{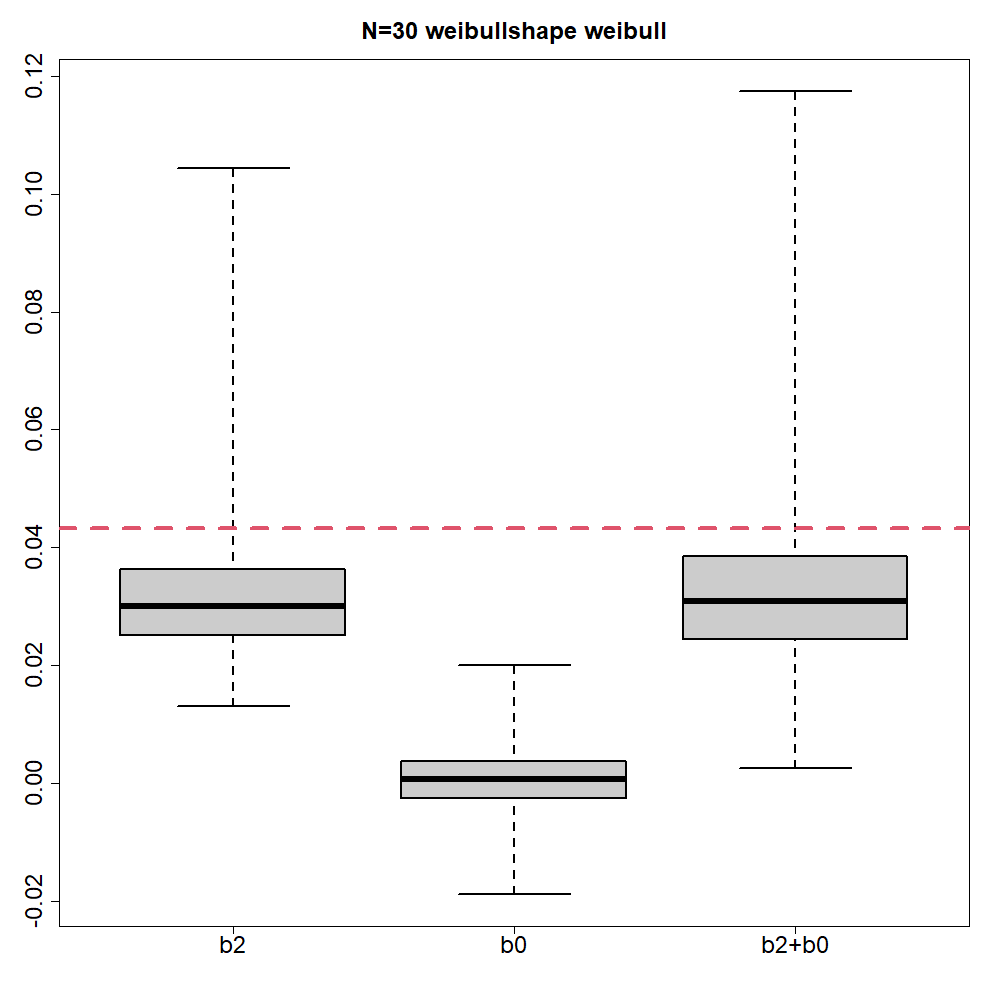}
    \caption{Bias estimation of the parameters in the Weibull fitting. The first and second panel correspond to the scale parameter $\lambda$ and shape parameter $\gamma$, respectively. In each panel, from left to right, results for $\hat{b}_2$, $\hat{b}_0$, and $\hat{b}_0+\hat{b}_2$ are shown as boxplots; each boxplot consists of the results from 5000 sets of artificial data. The horizontal dotted line colored red indicates the true value of the bias estimated from the average over 5000 sets of artificial data. 
    }
    \label{fig:weibull}
\end{figure}

The results are presented in Fig.\ref{fig:weibull}. The first and second panel of Fig.\ref{fig:weibull} correspond to the scale parameter $\lambda$ and shape parameter $\gamma$, respectively. In each panel, from left to right, results for $\hat{b}_2$, $\hat{b}_0$, and $\hat{b}_0+\hat{b}_2$ are shown. The horizontal dotted line colored red indicates the true value of the bias estimated from the average over the sets of artificial data. 

Fig.\ref{fig:weibull} indicates that the bias is mostly explained by $\hat{b}_0$ in the case of the scale parameter $\lambda$, while it is predominantly due to $\hat{b}_2$ in the case of the shape parameter $\gamma$. This exhibit the importance of both $\hat{b}_0$ and $\hat{b}_2$ in the estimation of the bias. Fig.\ref{fig:weibull} also presents that the proposed estimator $\hat{b}_0+\hat{b}_2$ provides reasonable estimates in both cases, while a small but systematic deviation is observed in the case of shape parameter; it may be related to a highly nonlinear nature of the estimator of $\gamma$.

\section{Example: logistic regression}
\label{sec:example_logistic}

Here we consider the logistic regression in a moderately high-dimensional settings, where the algorithm 2 is effectively applied. Let us consider the  logistic regression for data $Y^n=(Y_i), \, Y_i \in \{0,1\}$ as:
\begin{align}
    p(Y^n;a)=\prod_{i=1}^n q_i^{Y_i}(1-q_i)^{1-Y_i}, \qquad 
    \log \frac{q_i}{1-q_i}= \sum_{s=1}^{N_p} a_s x_{is}, 
\end{align}
where $a_s, \, s=1,\cdots,N_p$ is regression coefficients and $x_{is}, \, s=1,\cdots,N_p$ are explanatory variables corresponding to $Y_i$. We consider the following settings: 
\begin{itemize}
    \item The model comprises $N_p=3m$ explanatory variables. Here, we consider the cases of $N_p=60$ and $N_p=21$.
    \item Artificial data of sample size $n$ are generated using the same model. The values of the regression coefficients $(a_i)$ are $10$, $-5$, and $0$ in each of the three groups of the size $m$, respectively.
    \item The value of the explanatory variables are randomly selected from the normal distribution of the variance $1/n$ and mean $0$. When we consider the effect of confounding between explanatory variables, the values of the explanatory variables are sampled from a multivariate normal distribution; off-diagonal elements of the covariance matrix are set as $\rho/n$.
    \item We select $\delta$ that defines a scale of increments of $\lambda_i$s as $0.2$ throughout the experiments; an adaptive choice of $\delta$ is importnat, as well as an adequate stopping criterion of the algorithm, which is left for the future study.
\end{itemize}

\begin{figure}[ht]
    \centering
    \includegraphics[width=150mm]{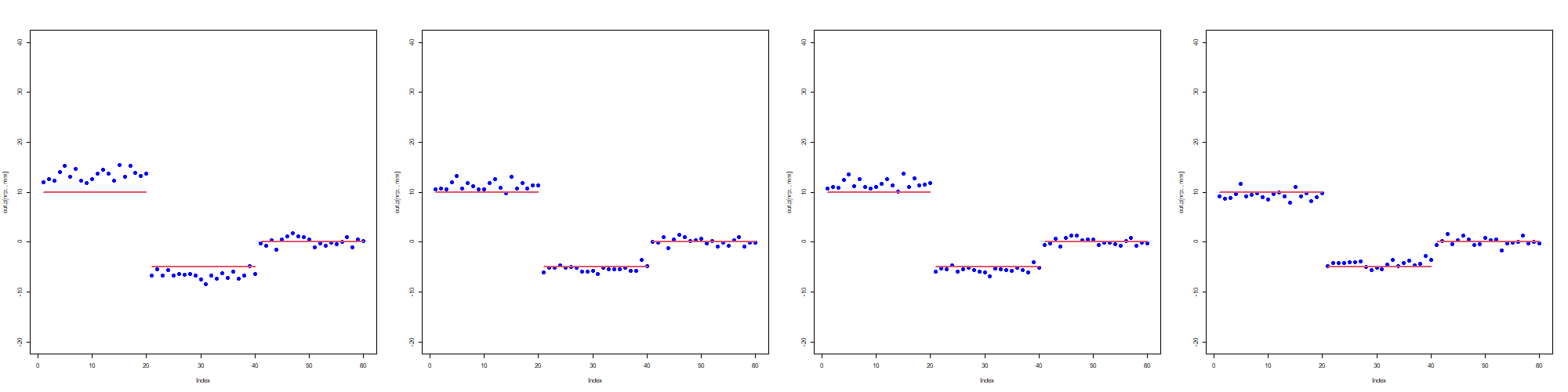}
    \caption{Bias correction in the logistic regression with $N_p=60$ and $n=600$. In each panel, we plot the median of the estimated regression coefficients $a_s, s = 1,\ldots, N_p$ in 30 trials. The true values of $a_s$ is expressed as horizontal red lines.  From left to right, the original estimates, estimates corrected by $\hat{b}_2$, $\hat{b}_0$, and $\hat{b}_0+\hat{b}_2$ are shown. 
    }
    \label{fig:alg1_good}
\end{figure}

\begin{figure}[ht]
    \centering
    \includegraphics[width=150mm]{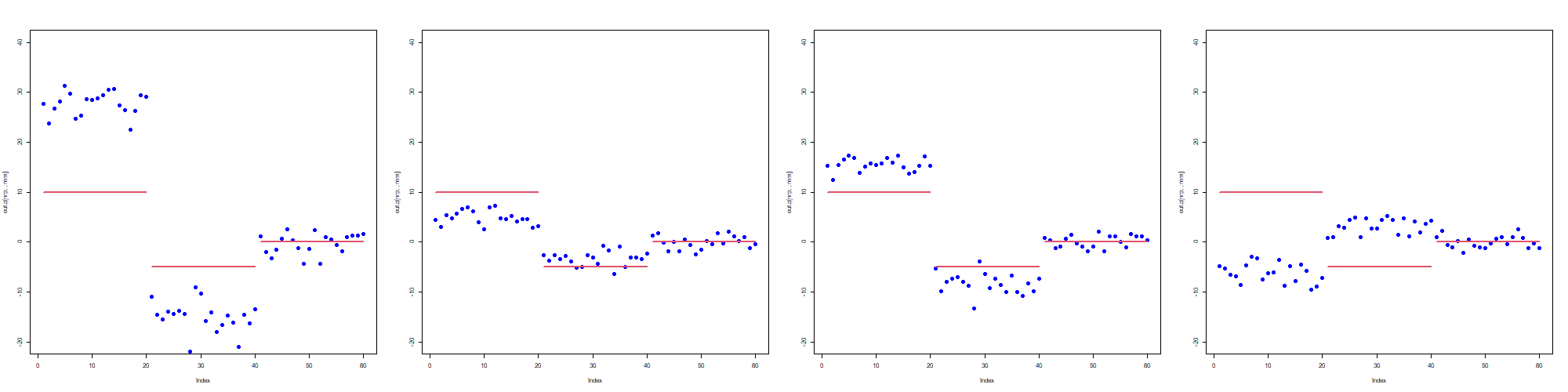}
    \centering
    \includegraphics[width=150mm]{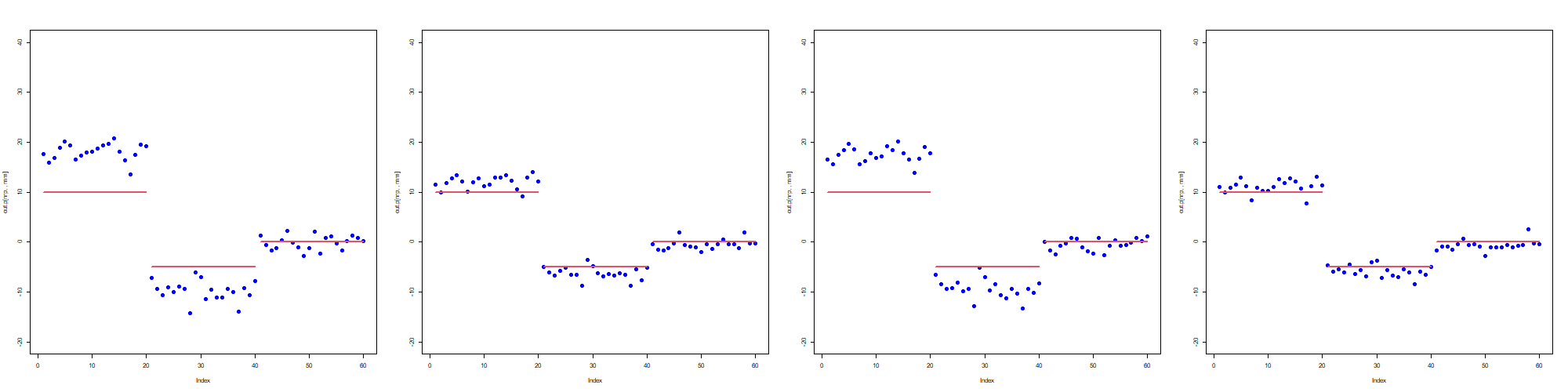}
    \caption{Bias correction in the logistic regression with $N_p=60$ and $n=300$. The first row presents the result using an original posterior with $\lambda_i=0$, while the second row presents the results after three iteration of updating $\lambda_i$s using \eqref{eq:iter_1} and \eqref{eq:iter_2}. Other details are the same as those in Fig.\ref{fig:alg1_good}.
    }
    \label{fig:alg2_good}
\end{figure}

\begin{figure}[ht]
    \centering
    \includegraphics[width=150mm]{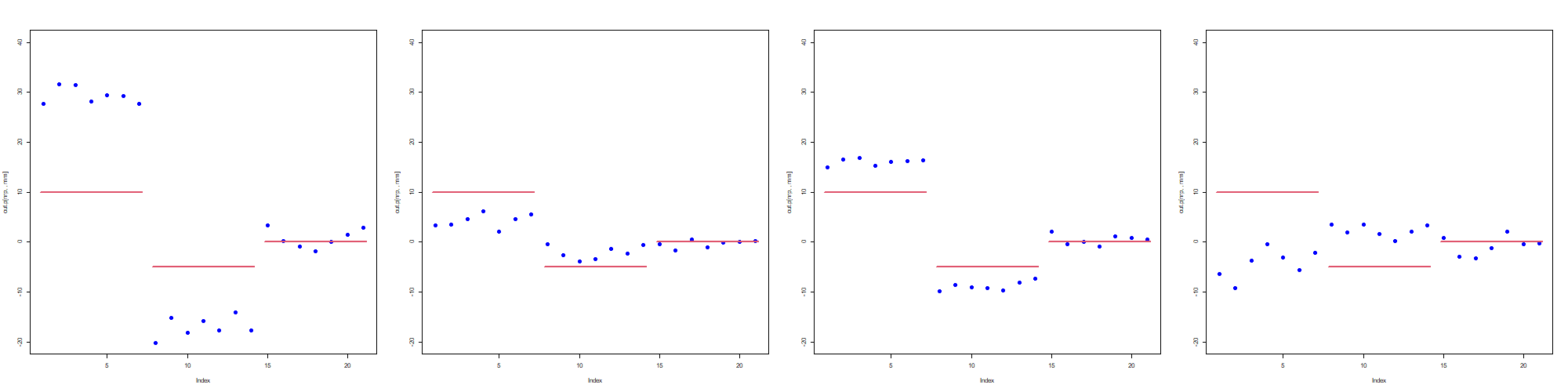}
    \centering
    \includegraphics[width=150mm]{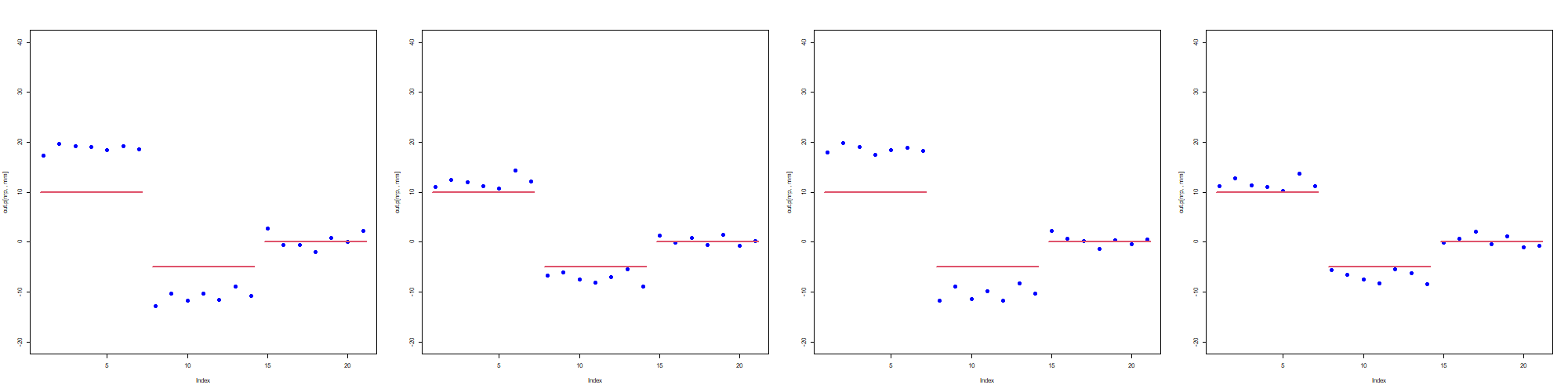}
    \caption{Bias correction in the logistic regression with confounded covariates. $N_p=21$ and $n=105$. The first row presents the result using an original posterior with $\lambda_i=0$, while the second row presents the results after three iteration of updating $\lambda_i$s using \eqref{eq:iter_1} and \eqref{eq:iter_2}. Other details are the same as those in Figs.\ref{fig:alg1_good} and \ref{fig:alg2_good}.
    }
    \label{fig:alg2_good2}
\end{figure}

In figures \ref{fig:alg1_good}--\ref{fig:alg2_good2}, medians of the estimated regression coefficients $a_s, \, s=1,\cdots, N_p$ in 30 trials are plotted; each of these 30 trials employs a different set of $Y_i$ and $x_{is}$. The true values of $a_s$ is expressed as horizontal red lines. 

Fig.~\ref{fig:alg1_good} presents a result of algorithm 1, where the number of parameters is $N_P=60$ and the sample size $n$ is set as $10\times N_p=600$. From left to right, the original estimates (posterior means), estimates corrected by $\hat{b}_2$, $\hat{b}_0$, and $\hat{b}_0+\hat{b}_2$ are presented; the rightmost panel corresponds to the output of algorithm 1. While the original estimates $\Eppos[a_s]$ in the leftmost panel have considerable bias, the algorithm 1 provides reasonably corrected values of $a_s$. In addition, results in the second and third panel indicate that both of $\hat{b}_2$ and $\hat{b}_0$ contribute the bias correction in algorithm 1.  

Fig.~\ref{fig:alg2_good} presents another result, where the sample size $n$ reduces to $5\times N_p=300$ with a corresponding change of the variance of explanatory variables. The first row of Fig.~\ref{fig:alg2_good} presents the results of the algorithm 1. In contrast to the case $n=600$, the rightmost panel indicates that the proposed estimator $\hat{b}_0 +\hat{b}_2$ fails to correct the bias. Results in the second and third panel indicate that adequate corrections are not achieved even when we select one of the terms $\hat{b}_0$ and $\hat{b}_2$ and discard the other. Numerical attempts of evaluating the accuracy of $\hat{b}_0$ and $\hat{b}_2$ separately (not shown here) suggest that the majority of the error in a sparse-data case comes from $\hat{b}_0$.

On the other hand, the second row of Fig.~\ref{fig:alg2_good} presents the result of algorithm 2 for the same sets of artificial data.  As shown in the rightmost panel shows, algorithm 2 successfully corrects the bias after only three iterations of updating $\lambda_i$s using \eqref{eq:iter_1} and \eqref{eq:iter_2}. We stress that the bias is not entirely removed even after updating $\lambda_i$s; it is evident in the leftmost panel in the second row, where the posterior means defined with the updated values of $\lambda_i$s are presented. This means that the role of the  bias correction terms $-\sum_{k=1}^K \lambda_k A_k(\theta)$ are to reduce the bias and make the bias correction by algorithm 1 effective.  

Thus far, we assumed that the values of explanatory variables $x_{is}$ are independently generated for each $s$; this assumption might hardly hold in a real-world problem. In Fig.\ref{fig:alg2_good2}, we present the results for artificial data of $N_p=21$ and $n=5\times Np=105$, where the values of the explanatory variables $x_{is}$ are correlated in the artificial sets of the data;  off-diagonal elements of the covariance matrix are set as $0.5/N_p$. The results are essentially the same as those in Fig.\ref{fig:alg2_good}: algorithm 1 fails to correct the bias as shown in the rightmost panel of the first row, while algorithm 2 successfully correct the bias as shown in the rightmost panel of the second row.

\section{Derivation}
\label{sec:derivation}

\subsection*{Settings and regularity conditions}

The posterior distribution $p_w(\theta ;X^n)$ with weights $(w_1,w_2,\ldots,w_n)$ of the observations $X^n=(X_1,X_2,\ldots X_n)$ is defined by: 
\begin{align}
    p_w(\theta ;X^n) = 
    \frac{\exp\{\sum_{i=1}^{n} w_i \ell(X_{i}; \theta)\}p(\theta) }
    {\int \exp\{\sum_{i=1}^{n} w_i \ell(X_{i};\theta')\}p(\theta') d\theta'}.
    \label{eq:weighted_pos}
\end{align}
The average of $A(\theta)$ over the distribution $p_w(\theta ;X^n)$ are expressed as $\Eppos^w[A(\theta)]$. Hereafter $w=1$ is used as an abbreviation of $w_i=1, i=1,\ldots,n$.

We also introduce the following notation for a fourth order combined cumulant: 
\begin{align}
\KposABBBw & [A(\theta),B(\theta)]  = \\
& \Eppos^w \left [(A(\theta)-\Eppos[A(\theta)])( B(\theta)-\Eppos^w [B(\theta)])^3 \right] 
\\
& -3 \times \Eppos^w \left [(A(\theta)-\Eppos^w[A(\theta)])(B(\theta)- \Eppos^w [B(\theta)]) \right ]
\times
\Eppos^w \left [(B(\theta)-\Eppos^w[B(\theta)])^2 \right ].
\label{eq:poscum_4} 
\end{align}

We assume the following conditions (C1) and (C2). Hereafter $\EpX[\,\,\,]$ denotes a population average over an observation $X_1$;  $\mathcal{W}^{(-1)}$ is the set defined by $0 \le w_1 \le 1$ and $w_{i\neq 1}=1$. 

\begin{enumerate}
    \item[(C1)] 
    $
        \EpXi\left[
        \sup_{w\in \mathcal{W}^{(-1)}}
        \left|
        \KposABBBw[A(\theta),\ell(X_1;\theta)]
        \right|
        \right]
         =o(1/n^2)  
    $
    \item[(C2)] There exists an integrable function $M(\theta)$ such that
    for all $w\in \mathcal{W}^{(-1)}$, 
    \begin{align}
    &\left| p(\theta)
    \exp\left(w_{1}\ell(X_{1};\theta)+
    \sum_{k\neq 1} \ell(X_{k};\theta) \right)
    A^l(\theta) \{\ell(X_1;\theta) \}^m \right| \,
    \le M(\theta), \,\,\, 0 \le l \le 1, \,\, 0 \le m \le 3.
    \end{align}
\end{enumerate}

\subsection*{Local case sensitivity formulae}

Let us introduce local case sensitivity formulae as a basic tool for the following derivation. The first-order local case sensitivity formula (\citealp{Gustafson_1996, Perez_etal_2006, MillarandStewart(2007)}) represents the first-order derivative of $\Eppos^w[A(\theta)]$ using the posterior covariance as
\begin{align}
& \left. \frac{\partial}{\partial w_i} \Eppos^w[A(\theta)] \right |_{w=1} 
=\Covpos[A(\theta),  \ell(X_{i};\theta)].
 \label{eq:local_1}
\shortintertext{In a similar manner, special cases of second- and third- order formulae required here are expressed as}
& \left. \frac{\partial^2}{\partial w_i^2} \Eppos^w[A(\theta)] \right |_{w=1}
= \Kpos[A(\theta), \ell(X_{i};\theta), \ell(X_{i};\theta)].
 \label{eq:local_2}
\\
& \frac{\partial^3}{\partial w_i^3} \Eppos^w[A(\theta)] 
= \KposABBBw[A(\theta), \ell(X_{i};\theta)],
 \label{eq:local_3}
\end{align}
where the last one is defined for an arbitrary $0 \le w \le 1$. The proof of these formulae is straightforward when an exchange of integration and derivation is allowed under the condition (C2); it is presented in \cite{Bayesian_IJK} in a generic form; \eqref{eq:local_2} is also proved in \cite{Iba_Yano_arXiv2}.

\subsection*{Bayesian IJK for the bias}

The algorithm 1 is derived as a Bayesian IJK approximation to the jackkinfe bias correction formula:
\begin{align}
\label{eq:jackknife_bias}
    b_{\mathrm{jack}}= (n-1) \left (\frac{1}{n} \sum_{i=1}^n \Eppos^{-i}[A] -\Eppos[A] \right )
    = \sum_{i=1}^n \left (\Eppos^{-i}[A] -\Eppos[A] \right ) +o_p(1/n).
\end{align}
This formula is well known, but we give a derivation in the appendix~\ref{app:jack}. In the jackknife method, the number of the observations are changed from $n$ to $n-1$ by the deletion of an observation.  It is essential to deal with the difinitional bias. Seemingly more sophisticated sample-reuse approaches, such as the bootstrap method, may fail in dealing with the definitional bias, when the number of the observations is kept constant before and after disturbing the data.

Setting $w_i=0$ and $w_j=1 \, (j\neq i)$ in the weighted posterior \eqref{eq:weighted_pos} and consider a Taylor expansion with respect to $w_i-1$ of the posterior expectation, we obtain
\begin{align}
\label{eq:delete_one}
\quad
    \Eppos^{-i}[A] = \Eppos[A]
    -\left. \frac{\partial}{\partial w_i} \Eppos^w[A(\theta)] \right |_{w=1}
    \!\!\!\!
    +\frac{1}{2} \left. \frac{\partial^2}{\partial w_i^2} \Eppos^w[A(\theta)] \right |_{w=1} 
    \!\!\!\!
    - \frac{1}{6} \sum_{i=1}^n \left. \frac{\partial^3}{\partial w_i^3} \Eppos^w[A(\theta)] \right |_{w=w*}
\end{align}
where $0 \le w* \le 1$. On the other hand, the expression \eqref{eq:local_3} and condition (C1) gives
\begin{align}
\label{eq:delete_one_op}
    \frac{1}{6} \sum_{i=1}^n \left. \frac{\partial^3}{\partial w_i^3} \Eppos^w[A(\theta)] \right |_{w=w*} =o_p(1/n).
\end{align}

Substituting \eqref{eq:delete_one} in \eqref{eq:jackknife_bias} and use \eqref{eq:delete_one_op}, $b_{\mathrm{jack}}$ is expressed as
\begin{align}
\label{eq:jack_Taylor}
    b_{\mathrm{jack}}= - \sum_{i=1}^n \left. \frac{\partial}{\partial w_i} \Eppos^w[A(\theta)] \right |_{w=1}
    +\frac{1}{2} \sum_{i=1}^n \left. \frac{\partial^2}{\partial w_i^2} \Eppos^w[A(\theta)] \right |_{w=1}
    + o_p(1/n).
\end{align}
Now that we use \eqref{eq:local_1}
and \eqref{eq:local_2} in the expansion \eqref{eq:jack_Taylor}, which completes the derivation as
\begin{align}
    b_{\mathrm{jack}}= - \sum_{i=1}^n \Covpos[A(\theta),  \ell(X_{i};\theta)]
    +\frac{1}{2} \sum_{i=1}^n \Kpos[A(\theta), \ell(X_{i};\theta), \ell(X_{i};\theta)]
    + o_p(1/n).
\end{align}

\subsection*{Derivation of the algorithm 2}

The algorithm 2 is considered as an iterative improvement of the residual bias based on the following representation of the derivative
\begin{align}
     \frac{\partial}{\partial \lambda_k} \Epposl[\theta_{k^\prime}]=
     -\Covposl[\theta_k, \theta_{k^\prime}],
     \label{eq:lambda_diff}
\end{align}
where $\Covposl[\,\,]$ denote the covariance with the distribution \eqref{eq:pos_lambda}. The expression \eqref{eq:lambda_diff} is obtained by a direct computation when an exchange of integration and derivation is allowed under appropriate regularity conditions. The idea behind \eqref{eq:lambda_diff} is the same as that of local case sensitivity formulae; applications of similar formulae have a long history in statistics (\cite{Geyer1993, Perez_etal_2006}). 

Let us define $\lambda^*$ as a value of $\lambda$ that gives zero bias for all parameters $\theta_k$; such a value might not exist, but here we introduce it as a hypothetical target of the bias reduction procedure.
\begin{align}
\hat{b}_k(\theta;n;\lambda^*)=0.
\label{eq:lambda_eq}
\end{align} 
If we define $\triangle\lambda^{(m)}=\lambda^*-\lambda^{(m)}$, where $\lambda^{(m)}$ is the value of $\lambda$ at step $m$.
Then, we expand \eqref{eq:lambda_eq} around $\lambda^{(m)}$ as
\begin{align}
    \hat{b}_{k}(\theta;n;\lambda^{(m)})
    +\sum_{k^\prime} \frac{\partial}{\partial \lambda_{k^\prime}} \Epposl[\theta_{k}]
    \triangle\lambda^{(m)}_{k^\prime}
    + o(\triangle\lambda^{(m)})=0.
\end{align} 
Using \eqref{eq:lambda_diff} and ignore the residual terms, this is expressed as
\begin{align}
    \sum_{k^\prime} \Covposl[\theta_k(\theta), \theta_{k^\prime}(\theta)]
    \triangle\lambda^{(m)}_{k^\prime}=\hat{b}_{k}(\theta;n;\lambda^{(m)}).
\end{align} 
In a vector form, this gives the equation \eqref{eq:iter_1}.

\section{Summary and future problems}

We proposed algorithms for estimating and correcting the bias of the posterior mean as an estimator of parameters. Two algorithms are introduced, both of which rely on posterior covariance and cumulants derived from MCMC outputs, eliminating the need for additional analytical calculations. The first algorithm is based on a Bayesian infinitesimal jackknife approximation and successfully estimate the bias including the difinitional bias (\cite{Efron_2015}) using the result of a single MCMC run. The second algorithm involves an iterative improvement of a quasi-prior based on the output of the first algorithm; it is shown to be effective in high-dimensional and sparse settings for logistic regression.   

In addition to utilizing the recently emerging Bayesian infinitesimal jackknife approximation (\cite{Bayesian_IJK}), the second algorithm is characterized by a hybrid approach that combines bias estimation and correction. An significant aspect of this strategy is that the ``bias reduction terms'' do not need to completely eliminate the bias. Instead, their role is to reduce the bias sufficiently so that the bias correction by the first algorithm becomes effective. However, the use of the first algorithm within the second algorithm requires justification of the first algorithm in cases where quasi-prior is non-negligible. Preliminary analysis suggests that the success of this approach may depend on the use of a quasi-prior whose logarithm is linear in $\theta$, but further analysis and experimentation are required for the understanding.

There are a number of subjects left for future studies: First, more theoretical analysis is needed along with the development of an appropriate stopping criterion for the algorithm. Additionally, the relationship between the proposed method and existing approaches, such as the Firth method for bias correction of the maximum likelihood estimator (\cite{Firth1993}), should be explored. Finally, applications to large-scale real-world data and further tests using artificial data are necessary to assess the potential and limitations of the proposed method.

\section*{Acknowledgements}

I would like to thank Keisuke Yano for the fruitful discussions and valuable advice.

\appendix

\section{von Mises expansion}
\label{app:von_Mises}

Here we introduce the von Mises expansion and provide an interpretation of the definitional bias $b_0$ in this context. Let us define a formal posterior for an arbitrary distribution $F$ as
\begin{align}
    p(\theta ; F,n ) = 
    \frac{\exp\{n \int \ell(x; \theta) dF(x) \}p(\theta) }
    {\int \exp\{n \int \ell(x; \theta') dF(x) \}p(\theta') d\theta'},
    \label{eq:F_pos}
\end{align}
where $F$ substitutes for the empirical distribution $\hat{G}_n$ in \eqref{eq:pos}; here as usual the empirical distribution is defined by the sum $d\hat{G}_n=(1/n)\sum_{i=1}^n \delta_{X_i}$ of the point measures $\delta_{X_i}$ concentrated on observations $X^n=(X_1,X_2, \ldots, X_n)$. 

An estimator $T^A(F, n)$ of $A(\theta)$ is defined as a posterior mean as
\begin{align}
T^A(F, n)=\int A(\theta)p(\theta;F,n) d\theta.
 \label{eq:av_F_pos}
\end{align} 
Specifically, $T^A(\hat{G}_n,n)=\Eppos[A(\theta)]$. When $F=G$, $T^A(G,n)$ defines an ``ideal value at sample size $n$'' for the estimator; this does not necessarily coincides with the ``true value'' (or projection) defined as $A_0=\lim _{n \rightarrow \infty}T^A(G,n)$ and can have some bias even for a consistent estimator.

Now that we introduce von Mises expansion (\cite{mises1947asymptotic, Konishi_Kitagawa_book, Bayesian_IJK}) as
\begin{align}
T^A(\hat{G},n)=T^A(G,n)+\frac{1}{n} \sum_{i=1}^n T^A_1(X_i;G)
+\frac{1}{2n^2} \sum_{i=1}^n \sum_{j=1}^n T^A_2(X_i,X_j;G)+o_p \left (\frac{1}{n} \right ),
 \label{eq:vonM}
\end{align}
where $T^A_1(X_i;G)$ and $T^A_2(X_i,X_j;G)$ are influence functions for the true distribution $G$. These influence functions are assumed to satisfy
\begin{align}
\EpX[T^A_1(X_i;G)]=0, \,\,\, \Ep_{X_i}[T^A_2(X_i,X_j;G,n)]=\Ep_{X_j}[T^A_2(X_i,X_j;G,n)]=0,
 \label{eq:vonM_ave0}
\end{align}
where $\Ep_{X_i}$ means frequentist expectation over $X_i$, keeping other components of $X^n$ fixed.  
The dependence of $T^A_1$ and $T^A_2$ on the sample size $n$ is omitted here; we will find that it is crucial in the first term $T^A(G,n)$ for our purpose of estimating the bias. If we take the frequentist expectation for both sides and  use \eqref{eq:vonM_ave0}, it gives:
\begin{align}
\EpX[T^A(\hat{G},n)]= \EpX[T^A(G,n)]+\frac{1}{2n^2}\sum_{i=1}^n \EpX[T^A_2(X_i,X_i;G)]+o \left (\frac{1}{n} \right ).
 \label{eq:vonM_expect}
\end{align}

If we restrict ourselves to a consistent estimator and regard $\lim_{n \rightarrow \infty} T^A(G,n)$ as the true value of $A$, the bias of the estimator $T^A(\hat{G},n)$ is given by:
\begin{align}
&  b_0(A,n)=\EpX[T^A(G,n)]-\lim_{n \rightarrow \infty} T^A(G,n) \label{eq:vonM_bias0}, \\
& b_2(A,n)=\frac{1}{2n^2}\sum_{i=1}^n \EpX[T^A_2(X_i,X_i;G)],
 \label{eq:vonM_bias2} \\
 & b(A,n)=  \EpX[T^A(\hat{G},n)]-\lim_{n \rightarrow \infty} T^A(G,n)=b_0(A,n)+b_2(A,n)+o \left (\frac{1}{n} \right ), \label{eq:vonM_bias}
\end{align}
where $b_0(A,n)$, $b_2(A,n)$, and $b(A,n)$ may be abbreviated as $b_0$, $b_2$, and $b$, respectively.

It is natural to estimate $b_2$ using $\hat{b}_2$ defined by \eqref{eq:b2}, because the formula \eqref{eq:local_2} indicates that the second-order influence function $T^A_2(X_i,X_i;G)$ can be represented by the third-order posterior cumulant. To be precise, we need to consider relations in \eqref{eq:vonM_ave0},
\begin{align}
\Ep_{X_i}[T^A_2(X_i,X_j;G,n)]=\Ep_{X_j}[T^A_2(X_i,X_j;G,n)]=0,
\end{align}
imposed on the second order-influence function. However, the corresponding correction appears to have little effect in examples. If we consider $\hat{b}_2$ as an estimator of $b_2$, the rest part $\hat{b}_0$ of the proposed estimator should be regarded as an estimator of $b_0$; we can confirm this in an example of the binomial likelihood. 

\section{Jackknife bias correction}
\label{app:jack}

First, we assume the bias $b$ of the estimator $\hat{A}(X^n)$ of statistics $A$ is asymptotically proportional to the inverse of the sample size $n$ as:
\begin{align}
    \EpX[\hat{A}(X^n)]=A_0+\frac{c}{n}+r(n), 
\label{eq:n}
\end{align}
where $A_0$ is a ``true value'' of $A_0$ defined as a limit $\lim_{n\rightarrow \infty} \hat{A}(X^n)$; $c$ is a constant and the residual term $r(n)$ is assumed to be in the order of $O(1/n^{1+\alpha}), \alpha>0$.

Let us consider modified data where the observation $i$ is removed and express it as $\hat{A}(X^{-i})$. Since sample size of $X^{-i}$ is $n-1$ and each component comes from the same population as for $X^n$, \eqref{eq:n} indicates
\begin{align}
    \EpX[\hat{A}(X^{-i})]=A_0+\frac{c}{n-1}+r(n-1).
    \label{eq:n-1}
\end{align}
From \eqref{eq:n} and \eqref{eq:n-1}, we obtain the desired result as 
\begin{align}
    \EpX\left [ \sum_{i=1}^n \bigg (\hat{A}(X^{-i}) - \hat{A}(X) \bigg ) \right ]=n  \left (\frac{c}{n-1}-\frac{c}{n} \right ) +o(1/n)
    = \frac{c}{n}+o(1/n).
\end{align}

\clearpage
\bibliography{iba}

\begin{thebibliography}{}

\bibitem[Efron, 2015]{Efron_2015}
Efron, B. (2015).
\newblock Frequentist accuracy of {Bayesian} estimates.
\newblock {\em Journal of the Royal Statistical Society. Series B}, 77:617--646.

\bibitem[Firth, 1993]{Firth1993}
Firth, D. (1993).
\newblock Bias reduction of maximum likelihood estimates.
\newblock {\em Biometrika}, 80(1):27--38.

\bibitem[Geyer and Thompson, 1992]{Geyer1993}
Geyer, C.~J. and Thompson, E.~A. (1992).
\newblock Constrained {Monte Carlo} maximum likelihood for dependent data.
\newblock {\em Journal of the Royal Statistical Society. Series B}, 54(3):657--699.

\bibitem[Giordano and Broderick, 2023]{Bayesian_IJK}
Giordano, R. and Broderick, T. (2023).
\newblock The {Bayesian} infinitesimal jackknife for variance.
\newblock arXiv:2305.06466.

\bibitem[Gustafson, 1996]{Gustafson_1996}
Gustafson, P. (1996).
\newblock Local sensitivity of posterior expectations.
\newblock {\em The Annals of Statistics}, 24:174--195.

\bibitem[Iba and Yano, 2022]{Iba_Yano_arXiv2}
Iba, Y. and Yano, K. (2022).
\newblock Posterior covariance information criterion for arbitrary loss functions.
\newblock arXiv:2206.05887.

\bibitem[Konishi and Kitagawa, 2008]{Konishi_Kitagawa_book}
Konishi, S. and Kitagawa, G. (2008).
\newblock {\em Information Criteria and Statistical Modeling}.
\newblock Springer.

\bibitem[Millar and Stewart, 2007]{MillarandStewart(2007)}
Millar, R. and Stewart, W. (2007).
\newblock Assessment of locally influential observations in {Bayesian} models.
\newblock {\em Bayesian Analysis}, 2:365--384.

\bibitem[P\'{e}rez et~al., 2006]{Perez_etal_2006}
P\'{e}rez, C., Martin, J., and Rufo, M. (2006).
\newblock {MCMC}-based local parametric sensitivity estimations.
\newblock {\em Computational Statistics \& Data Analysis}, 51:823--835.

\bibitem[von Mises, 1947]{mises1947asymptotic}
von Mises, R. (1947).
\newblock On the asymptotic distribution of differentiable statistical functions.
\newblock {\em The Annals of Mathematical Statistics}, 18:309--348.

\end{thebibliography}

\bibliographystyle{apalike}

\end{document}